\documentclass[sigconf]{acmart}

\usepackage{booktabs} 
\usepackage{balance}
\usepackage{filecontents}
\usepackage{xurl}
\usepackage{xcolor}
\usepackage{subfig}
\usepackage{tabularx}

\newcolumntype{Y}{>{\centering\arraybackslash}X}

\usepackage[justification=centering]{caption}

\usepackage{enumerate}
\usepackage{enumitem}
\usepackage[normalem]{ulem}
\usepackage{float}
\usepackage{multicol}
\usepackage{multirow}

\usepackage{tabularx}

\begin{document}
\balance
\title[COVID-19 Misinformation Warning Labels]{Misinformation Warning Labels: Twitter's Soft Moderation Effects on COVID-19 Vaccine Belief Echoes}

\author{Filipo Sharevski}
\affiliation{%
  \institution{DePaul University}
  \streetaddress{243 S Wabash Ave}
  \city{Chicago}
  \state{IL}
  \postcode{60604}
}
\email{fsharevs@cdm.depaul.edu}

\author{Raniem Alsaadi}
\affiliation{%
  \institution{DePaul University}
  \streetaddress{243 S Wabash Ave}
  \city{Chicago}
  \state{IL}
  \postcode{60604}
}
\email{ralsaad2@depaul.edu}

\author{Peter Jachim}
\affiliation{%
  \institution{DePaul University}
  \streetaddress{243 S Wabash Ave}
  \city{Chicago}
  \state{IL}
  \postcode{60604}
}
\email{pjachim@depaul.edu}

\author{Emma Pieroni}
\affiliation{%
  \institution{DePaul University}
  \streetaddress{243 S Wabash Ave}
  \city{Chicago}
  \state{IL}
  \postcode{60604}
}
\email{epieroni@depaul.edu}

 \renewcommand{\shortauthors}{F. Sharevski, R. Alsaadi, P. Jachim, E. Pieroni}

\begin{abstract}
Twitter, prompted by the rapid spread of alternative narratives, started actively warning users about the spread of COVID-19 misinformation. This form of soft moderation comes in two forms: as a warning cover before the Tweet is displayed to the user and as a warning tag below the Tweet. This study investigates how each of the soft moderation forms affects the perceived accuracy of COVID-19 vaccine misinformation on Twitter. The results suggest that the warning covers work, but not the tags, in reducing the perception of accuracy of COVID-19 vaccine misinformation on Twitter. ``Belief echoes'' do exist among Twitter users, unfettered by any warning labels, in relationship to the perceived safety and efficacy of the COVID-19 vaccine as well as the vaccination hesitancy for themselves and their children. The implications of these results are discussed in the context of usable security affordances for combating misinformation on social media.     
\end{abstract}



\maketitle

\vspace{3em}

\section{Introduction}
In 2016, when ``fake news'' gained enormous popularity, Facebook started adding tags that say ``disputed'' on stories that were debunked by fact-checkers \cite{Mosseri}. About a year later, Facebook started adding fact-checks under potentially misleading stories \cite{Smith}. The goal of these initiatives was presumably to minimize the probability that readers will believe the fake information. Twitter did not begin similar initiatives until 2020, when, in late March, the platform began issuing labels on Tweets deemed as spreading misinformation related to the COVID-19 pandemic \cite{Roth}. According to Twitter, they are relying on their team and internal systems to monitor COVID-19 content for false or misleading information that is not corroborated by public health authorities or subject matter experts. The supposed aim of these labels is to reduce misleading or harmful information that could ``incite people to action and cause widespread panic, social unrest or large-scale disorder'' \cite{Roth}. Originally, only Tweets that pertained to COVID-19 were flagged; however, following the November 2020 presidential election, Twitter broadened the types of misleading, false or disputed information to which it appended warning labels about the outcome of the election, claims of election fraud, or the safety of voting by mail \cite{Spangler}. 

However, there is no evidence that these labels are effective, and in fact, an early investigation suggests that they can have an effect of ``backfiring'' i.e. convince people to believe the misinformation even more than if the label were not there \cite{Clayton}. A study exploring this so-called ``soft moderation'' implemented by Twitter found that the platforms' content with warning labels generated more action than content without said labels \cite{Zannettou}. The soft moderated Tweets were more likely to be distributed than a ``valid'' Tweet through discourse, not always because the soft moderated Tweets contained misinformation but because they contained a response to mock or disclaim an original author or a valid Tweet. Interestingly, a mere 1\% of the Tweets gathered for the study were labeled with a COVID-19 warning and a number of these few Tweets were found to be mislabeled simply because they contained the words ``oxygen'' and ``frequency''. Another study, in this context, found that some users did not trust the soft moderation intervention and felt that Twitter itself was biased and mislabeling content \cite {Geeng}. Even without the warning labels, a varying degree of users’ perceptions and beliefs regarding vaccines in general and about COVID-19 vaccines in particular plays a role in individuals' reaction to (mis)information Tweets. A study on vaccine misinformation spreading on social media platforms showed the effectiveness of ``influential users'' within an ``echo chamber'' of vaccine-fearing followers to be very high ~\cite{Featherstone20}. 

In this study, we set to examine the association between the beliefs regarding COVID-19 vaccines and the misinformation warning labels. To assess individuals’ beliefs about the COVID-19 vaccines, we used the questionnaire about COVID-19 vaccines from \cite{Biasio} adopted to the current state of the vaccine development and distribution (we conducted the study in January-February 2021 period, after the questionnaire was published and the vaccine was approved). We sought to examine the effect of different types of warning labels, represented as warning covers over the content or warning tags under the content \cite{Roth}, as experiments have shown that the design of the warning label itself (how the warning is presented to users) can affect the effectiveness of the warning \cite{Kim}. To examine the perceived accuracy of (mis)information Tweets about COVID-19 vaccine we utilized content from Twitter and a software that altered the warning tags to induce misperception. This software was initially developed to study polarized topics of discourse on social media on vaccines' safety and users' perception of accuracy contrary to their views and beliefs ~\cite{twittermim}. 

We found that the warning tags are ineffective in reducing the ``belief echoes'' on Twitter regarding the COVID-19 vaccine. The intended reduction effect, our results suggest, is achieved only with the misinformation warning covers preceding a misleading information (in our case, we use a Tweet referring to unverified adverse effects of the COVID-19 vaccination). We found that the less the users believed the COVID-19 vaccines are safe and efficacious, the more they perceived the misleading Tweets as accurate, even in presence of a warning tag below the Tweet's content. We also observed an echoing scepticism where the more the participants believed the COVID-19 vaccines are safe and efficacious, the less they perceived a Tweet being accurate, even in the case when they were presented with a verified COVID-19 vaccine information (A Tweet following Centers For Disease Control (CDC) guidelines in case there are any adverse effects after receiving the first COVID-19 vaccine dose). A similar echoing of beliefs and sceptic outlook of misleading and verified content, respectively, was found about the beliefs that herd immunity is a better option of immunization than mass COVID-19 vaccination. When it came to vaccine hesitancy, the warning labels did little to sway the participants on the benefits of the vaccination - the ones that were hesitant to receive the COVID-19 vaccine were convinced that it causes adverse effects leading to death, even if warned against such a claim. The anti-COVID-19 vaccine sentiment persisted, only in the case of the warning tags, when we asked whether children should get the vaccine too. 

We additionally explored the political affiliation and the effect of soft moderation, following the previously observed divisions on misinformation content along the main party lines \cite{Featherstone20, Zannettou}. We found that the Republicans and independent participants perceived the misleading Tweet as ``somewhat accurate'' while the Democrat participants perceived it as ``not very accurate,'' regardless of the presence of a warning tag. We also found that almost one in four Republicans and and one in six independent participants didn't expect to have a efficacious COVID-19 vaccine, while that proportion for the Democrats was one in forty. Half the Republicans and a third of the independent participants were hesitant to receive the COVID-19 vaccine, while only $1/20^{th}$ of the Democrats won't proceed with personal immunization. Roughly 40\% of the Republicans and independent participants were hesitant to vaccinate children for COVID-19, to which only 8.3\% of the Democrats agree with.  

We consider the misinformation warning labels as a form of usable security warnings akin to warnings about potentially harmful websites or favicons indicating unverified certificates \cite{Garfinkel}. Users, studies have shown, are reluctant to heed these warnings due to a lack of attention or motivation, incomprehension, or habituation \cite{Vance, Nicholson, Fagan}. The warning labels are written in plain language to draw attention to the user about the validity of the content. While it is early to assess the habituation effect of the warnings, the existence of the belief echoes posits an analysis of the unique blend of motivation and polarized habituation on Twitter \cite{Massachs}. We discuss the implications of our results in context of future designs of warning labels, as a form of a usable security interventions, aimed to curtail misinformation on social media.

\section{Background}
\subsection{Soft Moderation}
The misinformation warning labels (or tags) provide a compromise between content removal and the commitment of the social media platforms to allow for free and constructive discourse. However, whether such warnings and corrections/fact checks are effective in achieving their aim remains unclear. When measured based on Tweet engagement (e.g., likes, retweets or quote retweets), it appears that such warnings may be somewhat effective: Twitter reported a 29 \% decrease in quote Tweets that were labelled as misleading or disputed \cite{Spangler}. However, authors in \cite{Zannettou} found that Tweets with warning labels received more engagement (likes, retweets, replies, and quote retweets) than other Tweets from the same users that did not have warning labels. Specifically, \cite{Zannettou} found that between $2/3$ to $4/5^{th}$ of users receive more engagement on Tweets that contain content warning tags than Tweets that do not. Yet engagement is just one way to assess whether such warnings or corrections have an effect. This is especially the case since users’ engagement varied such that some users reinforced false claims, mocked the false claims, or debunked the claims \cite{Zannettou}. Thus, evaluating Tweet engagement alone is insufficient to evaluate whether content warning tags work to decrease the spread (or beliefs) of false information. 

Examining whether correcting or fact checking false information affects readers’ opinions and beliefs about the information, authors in \cite{Nyhan} found that correcting mock news articles that included false claims for politicians often failed to reduce misperceptions among particular ideological groups. The  corrections in studies often backfired, increasing misperceptions in the targeted group. The existence of so-called ``backfire effect'' suggests that providing corrections and fact checking information may have a counter-effect in combating false information (although the effect in \cite{Nyhan} is observed for mock news articles, not Facebook/Twitter posts). The backfiring effect was observed also in \cite{Zannettou} for the Elections 2020, which found that 72\% of the Tweets with warning labels were shared by Republicans while only 11\% are shared by Democrats. 

\subsection{Belief Echoes on Social Media}
More recently, studies have begun examining whether adding labels to posts on social media (as opposed to in articles) can affect individuals’ beliefs. For example, authors in \cite{Clayton} examined whether strategies that social media companies such as Twitter and Facebook use to oppose false stories or ``fake news'' would have the intended effect. This study also evaluated the efficacy of different types of labels: (i) a general warning, and (ii) two specific warnings pertaining to the article content. The authors found that a general warning had the intended effect of decreasing the perceived accuracy of the information but that adding ``disputed'' or ``rated false'' tags had a larger effect on minimizing perceived accuracy of the content, with the ``rated false'' tag most effective. Interestingly, and somewhat in contrast to the findings by Twitter regarding Tweet engagement \cite{Spangler}, the authors in \cite{Clayton} found that the tags did not reduce participants' self-reported likelihood of sharing the headlines on social media. Authors in \cite{Christenson} evaluated whether social media corrections of presidential Tweets on support of executive policies affects individuals’ attitudes. The idea was to test whether corrections are effective at rebutting false claims or whether they promote belief in the false claims among a particular demographic, a phenomenon dubbed as ``belief echoes'' \cite{Thorson}. The corrections had the intended effect on Democrats but the opposite effect on Republicans, showing evidence of the ``belief echoes'' on Twitter for the later category.  

The ''belief echoes'' manifest on social media when the exposure to negative political information continues to shape attitudes even after the information has been effectively discredited. Belief echoes can result as a spontaneous affective response that is immediately integrated into a person’s summary evaluation of social media content. The mere exposure to misinformation often generates a strong and automatic affective response, but the correction may not generate a response of an equal and opposite magnitude \cite{Gawronski}. One reason for this is that warning labels as commonly phrased as negations or contain exclamation marks. To combat the affectively asymmetrical soft moderation, users need to engage in cognitively demanding and time consuming ''strategic retrieval monitoring'' \cite{Ecker} or recall of the warning label. That does not happen often, so the misinformation may continue to affect evaluations, thus creating automatic belief echoes. Even if a person recalls the correction, they may discard it because they are already negatively predisposed to it. For example, in the context of politics, if a person hears that a candidate was accused of fraud, they may reason that the accusation emerged because the candidate is generally untrustworthy or corrupt. If these secondary inferences linger after the initial information is discredited, they will continue to affect their evaluations ~\cite{Thorson}.


\subsection{COVID-19 Vaccine Echo Beliefs}
Social media provides a vehicle for the spread of information regarding vaccines and vaccinations. Studies have found that most information on Twitter regarding vaccines is polarizing on the vaccine hesitancy and the beliefs about vaccines effects on child development \cite{Keim-Malpass}. The consumption of this information may affect individuals’ perceptions, attitudes and beliefs about vaccinations \cite{Massey}. For instance, vaccine-related Tweets by bots and trolls affect vaccine discourse on Twitter by promoting a relationship between vaccines and autism in children \cite{Broniatowski} or a relationship between COVID-19 vaccines and significant adverse effects, including death, for adults \cite{Allem}. Thus, misinformation regarding vaccines can have a significant effect on the acceptance of COVID-19 vaccines \cite{Cornwall}.

The current global pandemic provides ample opportunities for rampant misinformation regarding vaccines \cite{Vanderpool, EuropeanParliament}. This is particularly worrying because uptake of COVID-19 vaccines is critical for containing the spread of this disease and decreasing the morbidity and mortality imposed by the pandemic \cite{Lazarus}. Ensuring that individuals perceive the COVID-19 vaccines as safe once they become available requires that they have the correct information regarding COVID-19 vaccines \cite{Lazarus}. Currently, a significant minority of the worldwide population expresses skepticism about the safety, efficacy, and necessity of COVID-19 vaccines, which may make them more hesitant to take the COVID-19 vaccine. For instance, in Canada and the United States, 68.7\% and 75.3\%, respectively, reported being likely or very likely to accept the COVID-19 vaccine \cite{Lazarus}. Given the spread of the COVID-19 pandemic and the spread of misinformation regarding COVID-19 vaccines on Twitter \cite{Vanderpool}, it is imperative to explore the role of misinformation warning labels, as a form of usable security warnings, to curb misinformation pertaining to COVID-19 vaccines and vaccination more broadly.

\section{Research Study}
\subsection{Belief Echoes: Preconditions}
In this study, we set to examine the association between COVID-19 vaccine perceptions, beliefs, and hesitancy, the effect of the misinformation warning labels, and the perceived accuracy of (mis)information Tweets about COVID-19 vaccine content. First, we set to examine the \textit{preconditions} for existence of belief echoes on Twitter regarding COVID-19 vaccines. In particular, we investigated whether exposure to (mis)information Tweets about the COVID-19 vaccine efficacy in the presence or absence of warning labels, both as tags and covers, affect individuals’ perceptions of the Tweet’s accuracy with the following set of hypotheses: 

\begin{itemize}
\itemsep 0.5em
    \item H1: The presence of a warning tag under a Tweet containing \textit{misleading} information about COVID-19 vaccines will not reduce the perceived accuracy of the Tweet's content relative to a no warning tag condition. 

    \item H2: The presence of a warning cover before a Tweet containing \textit{misleading} information about COVID-19 vaccines is shown to the user will not reduce the perceived accuracy of the Tweet's content relative to a no warning cover condition.
        
    \item H3: The presence of a warning tag under a Tweet containing \textit{verified} information about COVID-19 vaccines will not reduce the perceived accuracy of the Tweet's content relative to a normal no warning tag condition. 
        
    \item H4: The presence of a warning cover (malware inserted) before a Tweet containing \textit{verified} information about COVID-19 vaccines is shown to the user will not reduce the perceived accuracy of the Tweet's content relative to a normal no warning cover condition.
\end{itemize}

To test the first hypothesis we utilized the Tweets containing \textit{misleading information} shown in Figure 1a and Figure 1b. The Tweet in Figure 1a shows a warning tag underneath a Tweet, indicating that the content is labeled as misinformation. The Tweet promulgates COVID-19 misinformation about a rare adverse effect that was linked to the SARS-CoV-2 virus, not the vaccine, at the time of writing \cite{Chappell}. To remove bias due to the ``influencer'' effect, the Tweet comes from a verified account named ``TheVaccinator'' (which we made up) and indicates a relatively high interaction engagement with 3k retweets, 13.5k quotations, and 12.8k likes, which is consistent with the expected engagement of Tweets containing COVID-19 vaccine information \cite{Zannettou}. An alteration of the same Tweet is shown in Figure 1b without the  accompanying warning tag. To test the second hypothesis we utilized the Tweets containing \textit{misleading information} shown in Figure 1b and Figure 2 (which includes a warning cover instead of a warning tag).

\begin{figure}[htb]
  \centering
  \includegraphics[width=\linewidth]{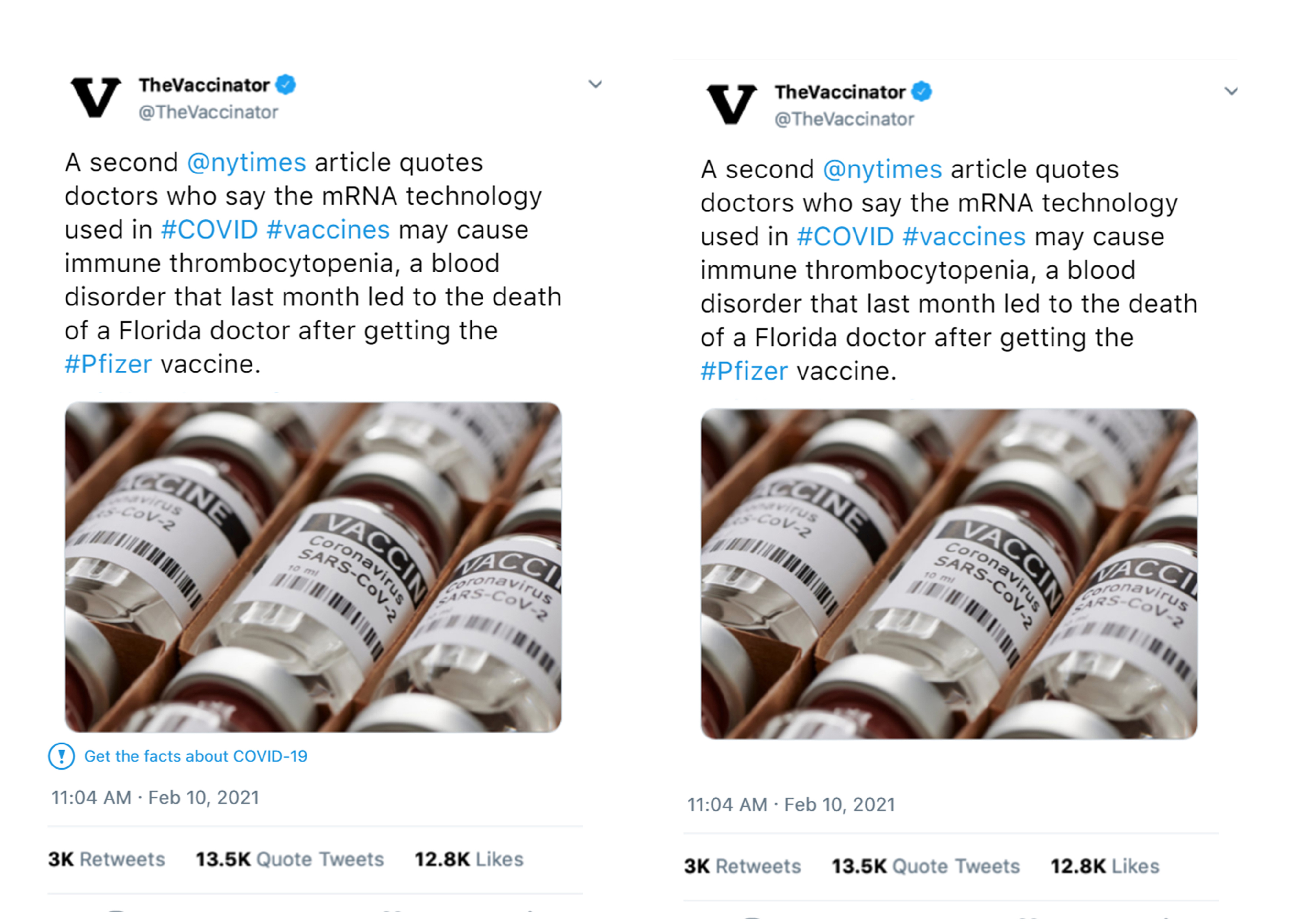}
  \subfloat[\label{MI with tag}]{\hspace{.5\linewidth}}
  \subfloat[\label{MI without tag}]{\hspace{.5\linewidth}}
  \caption[MI Tweet]{A Misleading Tweet: (a) \textit{With} a Warning Tag; \\(b) \textit{Without} a Warning Tag for Misleading Information.\label{MI tweet}}
\end{figure}

\begin{figure}[htb]
  \centering
  \includegraphics[width=\linewidth]{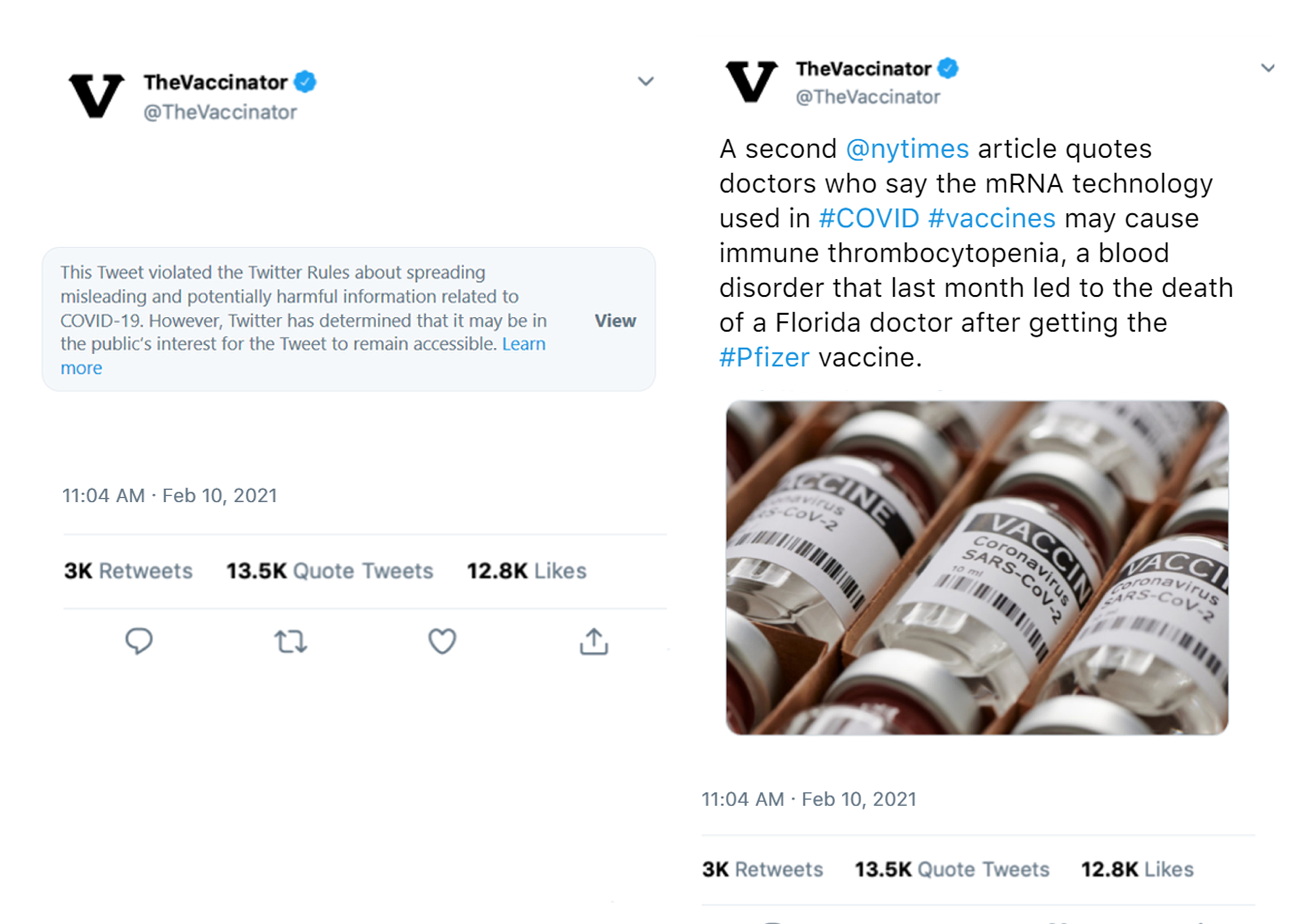}
  \caption{A Warning Cover Preceding a Misleading Tweet}
\end{figure}

To test the third hypothesis we utilized the Tweets containing \textit{verified information} shown in Figure 3a and Figure 3b. The Tweet content indicates the verified information distributed by the CDC about proceeding with the second dose of the COVID-19 vaccine in case an individual has a serious reaction from the first dose, altered to include a warning tag in Figure 3b \cite{CDC}. To control for bias, the Tweet comes from a verified account ``TheVirusMonitor'' instead of the CDC account and indicates a similar engagement as the misleading Tweet \cite{Zannettou}. To test the fourth hypothesis we utilized the Tweets containing \textit{verified information} shown in Figure 3a and Figure 4. We retained Figure 3a for the comparison of the conditions and altered the labeling in the Figure 3b to include a warning cover instead of a warning tag.

\begin{figure}[b]
  \centering
  \includegraphics[width=\linewidth]{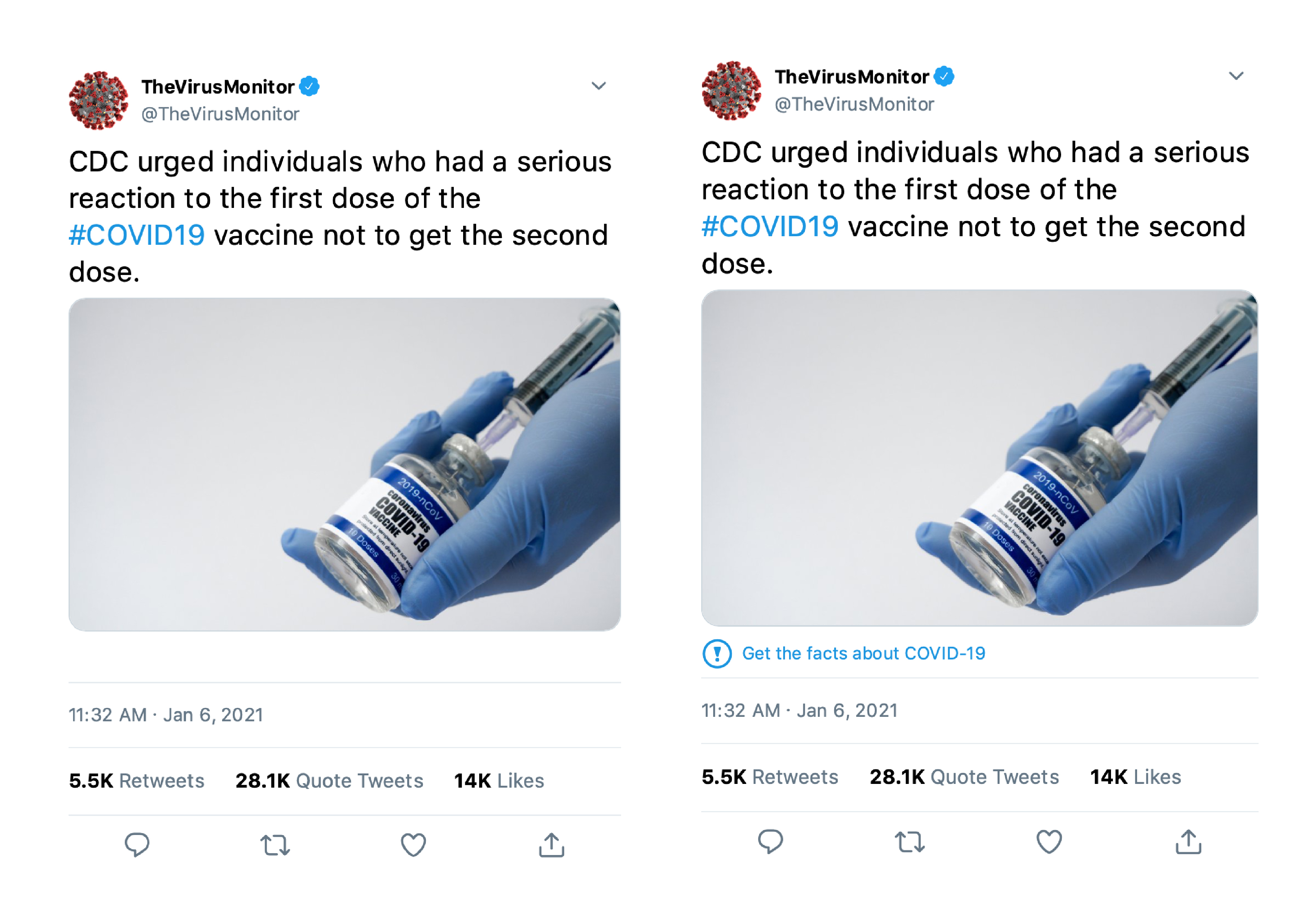}
  \subfloat[\label{VI without tag}]{\hspace{.5\linewidth}}
  \subfloat[\label{VI with tag}]{\hspace{.5\linewidth}}
  \caption[VI Tweet]{A Verified Tweet: (a) \textit{Without} a Warning Tag; \\ (b) \textit{With} a Warning Tag for Misleading Information.\label{VI tweet}}
\end{figure}

\begin{figure}[b]
  \centering
  \includegraphics[width=\linewidth]{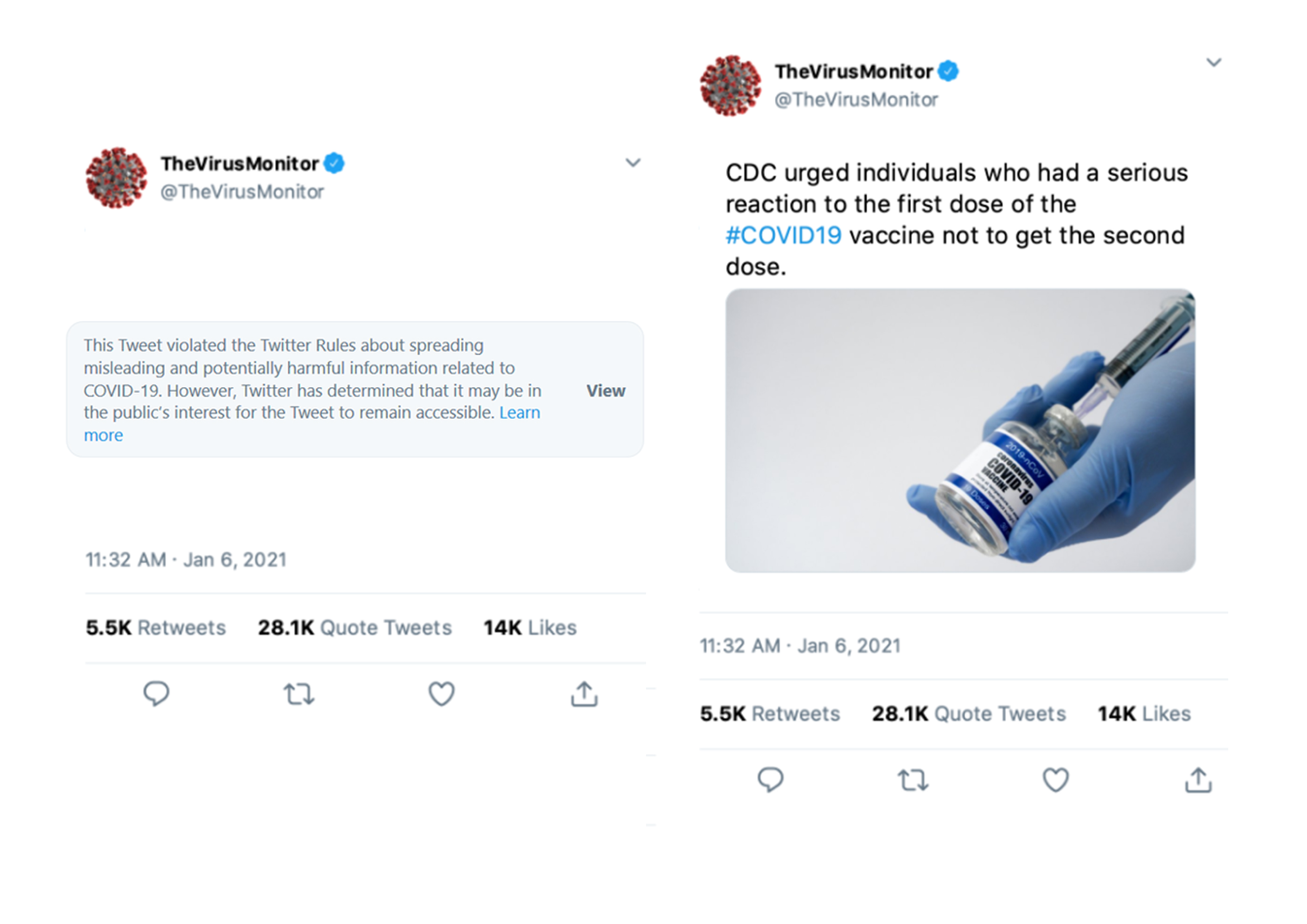}
  \caption{A Warning Cover Preceding a Verified Tweet}
\end{figure}

\subsection{Belief Echoes: Safety and Herd Immunity}
Assuming the preconditions of the belief echoes are met, we examined the relationship between COVID-19 vaccine beliefs on safety and herd immunity and the perceived accuracy of Tweets with COVID-19 vaccine information in presence/absence of warning labels. We used the same Tweets from Figures 1-4 together to test the following hypotheses:

\begin{itemize}
\itemsep 0.5em
        \item H5a: The belief that COVID-19 vaccines are not safe will not affect the perception of accuracy of a Tweet with misleading information about COVID-19 in any condition (with a warning tag/cover or without any warning)

        \item H5b: The belief that COVID-19 vaccines are not safe will not affect the perception of accuracy of a Tweet with verified information about COVID-19 in any condition (with a warning tag/cover or without any warning)

        \item H6a: The belief that there is no need for COVID-19 vaccine because herd immunity exists will not affect the perception of accuracy of a Tweet with misleading information about COVID-19 in any condition (with a warning tag/cover or without any warning)

        \item H6b: The belief that there is no need for COVID-19 vaccine because herd immunity exists will not affect the perception of accuracy of a Tweet with verified information about COVID-19 in any condition (with a warning tag/cover or without any warning)
       
\end{itemize}

\subsection{Belief Echoes: Efficacy and Hesitancy}
Next, we examined the relationship between COVID-19 vaccine efficacy/hesitancy and the perceived accuracy of Tweets with COVID-19 vaccine information in presence/absence of warning labels. We used the same Tweets as shown in Figures 1-4 together to test the following hypotheses:  

\begin{itemize}
\itemsep 0.5em
        \item H7a: The perception of producing efficacious COVID-19 vaccine will not affect the perception of accuracy of a Tweet with misleading information about COVID-19 in any condition (with a warning tag/cover or without any warning)

        \item H7b: The perception of producing efficacious COVID-19 vaccine will not affect the perception of accuracy of a Tweet with verified information about COVID-19 in any condition (with a warning tag/cover or without any warning)
        
        \item H8a: The COVID-19 vaccine personal hesitancy will not affect the perception of accuracy of a Tweet with misleading information about COVID-19 in any condition (with a warning tag/cover or without any warning)

        \item H8b: The COVID-19 vaccine personal hesitancy will not affect the perception of accuracy of a Tweet with verified information about COVID-19 in any condition (with a warning tag/cover or without any warning)

        \item H9a: The COVID-19 vaccine hesitancy for children will not affect the perception of accuracy of a Tweet with misleading information about COVID-19 in any condition (with a warning tag/cover or without any warning)

        \item H9b: The COVID-19 vaccine hesitancy for children will not affect the perception of accuracy of a Tweet with verified information about COVID-19 in any condition (with a warning tag/cover or without any warning)
\end{itemize}



\subsection{Belief Echoes and Political Affiliation}
To test the association between one's political affiliation and the perceived accuracy of the Tweets from Figures 1-4, following the evidence in\cite{Zannettou} about the interplay between political affiliation and misinformation warning labels, we asked: 

\begin{itemize}
\itemsep 0.5em
        \item RQ1: Is there a difference in the perceived accuracy of COVID-19 misleading/verified Tweets with warning labels (tags or covers) between Republican, Democrat, and Independent users?
        
        \item RQ2: Is there a difference between the beliefs and subjective attitudes of the Twitter users about the COVID-19 vaccine based on their political affiliation?
        
\end{itemize}

\subsection{Sampling and Instrumentation}
We first got approval from our Institutional Review Board (IRB) for an anonymous, non-full disclosure study. We set to sample a population of US residents using Amazon Mechanical Turk that is 18 years or above old, is a Twitter user, and has encountered at least one Tweet into their feed that relates to COVID-19 vaccines. There were both reputation and attention checks to prevent from bots and poor responses. The survey took between 5 and 10 minutes and the participants were compensates with the standard rate for participation. The study questionnaire, incorporating the instruments from \cite{Clayton, Biasio}, is provided in the \hyperref[sec:appendix]{Appendix}. We utilized a 2x3 between group experimental design where participants were randomized into one of six groups: (1) misleading Tweet with a warning tag; (2) misleading Tweet without a warning tag; (3) misleading Tweet with a warning cover; (4) verified Tweet; (5) verified Tweet with a warning tag; and (6) verified Tweet with a warning cover. 

After participation, the participants were debriefed and offered the option to revoke their answers. We crafted the content of the Tweets to be of relevance to the participants such that they meaningfully engage with the Tweet’s content (i.e., their responses are not arbitrary). We assumed participants understood the Twitter interface and metrics and were aware of the COVID-19 pandemic in general. However, we acknowledge that the level of interest regarding the COVID-19 vaccines could vary among the individual participants, affecting the extent to which their responses reflect their opinions.


\section{Study Results}
We conducted an online survey (N = 319) in January and February 2021. The power analysis conducted with $G^{*}$ Power 3.1 \cite{Faul} revealed that our sample was large enough to
yield valid results for both Wilcoxon–Mann–Whitney U-test comparing two groups and Pearson's correlation (minimum of 44 per group). There were 180 (56.4\%) males and 133 (41.7\%) females, with 6 participants (1.8\%) identifying as trans males, non-binary or preferring not to answer. The age brackets in the sample were distributed as follows: 13 (4.1\%) [18 - 24], 108 (33.9\%) [25 - 34], 98 (30.1\%) [35 - 44], 46 (14.1\%) [45 - 54], 37 (11.6\%) [55 - 64], 17 (5.3\%) [65 - 74], and 2 (.6\%) [75 - 84]. In terms of education, 4 (1.3\%) had less than high school, 24 (7.6\%) had a high school degree or equivalent, 54 (17.0\%) had some college but no degree, 43 (13.6\%) had a 2-year degree, 139 (43.8\%) had a 4-year college degree, and 53 (16.7\%) had a graduate or professional degree. Our sample, while balanced on the other demographics, was Democrat-leaning with 59 (18.5\%) Republicans, 157 (49.2\%) Democrats, and 102 (32\%) Independent participants.

\subsection{Belief Echoes: Preconditions}
The test of preconditions for formation of belief echoes on Twitter about the COVID-19 vaccine we first hypothesized that the presence of a warning tag under a Tweet containing \textit{misleading} information about COVID-19 vaccines will not reduce the perceived accuracy of the Tweet's content relative to a no warning tag condition. Indeed, the Mann-Whitney U test comparing the perceived accuracy between the participants exposed to the Tweet from Figure 1a and Figure 1b, respectively was insignificant, as shown in Table \ref{tab:H1-H4}. Confirming the H1 hypothesis, the results suggest that the warning tags didn't have the intended effect on reducing the perceived accuracy of a misleading COVID-19 vaccine information on Twitter. This proves the existence of preconditions for echoing one's belief irrespective of the warning tags.  

However, this is not the case with the warning covers used to label misleading information. The Mann-Whitney U test comparing the perceived accuracy between the participants exposed to the Tweet from Figure 1b and Figure 2 was significant, rejecting the H2 hypothesis. The participants exposed to the warning cover (Figure 2) reported, on average, that the Tweet was ``not very accurate'' while the ones exposed only to the Tweet content (Figure 1b) that the Tweet was ``somewhat accurate.'' The warning covers showed the intended effect of decreasing the perceived accuracy of misleading COVID-19 vaccine information on Twitter, dispelling one's echo beliefs for Tweets labeled with covers.

\begin{table}[H]
    \centering
    \caption{Preconditions Tests: Hypotheses H1 to H4}
    \label{tab:H1-H4}
 \begin{tabularx}{\linewidth}{|Y|Y|Y|}
  \hline
     
        & \textbf{U-test} & \textbf{Significance}  \\
  \hline 
  \textbf{H1} & $U = 1620.5$ & $p = .217$  \\\hline 
  \textbf{H2} & $U = 1841$ & $p = .004^{*}$ \\\hline
  \textbf{H3} & $U = 1124$ & $p = .063$ \\\hline
  \textbf{H4} & $U = 1123.5$ & $p = .087$ \\\hline
  \multicolumn{3}{|l|}{\textit{Significance Level: $\alpha = 0.05$}} \\\hline
 \end{tabularx}
\end{table}

As we suspected, the warning labels didn't reduce the verified COVID-19 information, regardless of whether a warning cover or warning tag was presented as a soft moderation intervention. The Mann-Whitney U tests comparing the perceived accuracy between the participants exposed to the Tweet from Figure 3b and Figure 3a was insignificant as was the comparison between the exposures between Figure 3a and Figure 4 (Table \ref{tab:H1-H4}). The retaining of H3 and H4 hypothesis suggest that participants critically discern the content of a seemingly valid Tweet, rather than the soft moderation labeling. Considering the previous reservations about the soft moderation implemented by Twitter when it comes to verified information \cite{Geeng}, these results suggest that preconditions for echoing one's beliefs exist irrespective of any warning labeling also in the case where these beliefs are rooted in verifiable facts about COVID-19.

\subsection{Belief Echoes: Safety and Herd Immunity}
With the evidence of existing preconditions of belief echoes about the COVID-19, both for misleading and verified information, we set to explore how these belief echoes materialize.  We asked the participants to what extent do they agree with the following statement: ``I am not favorable to the COVID-19 vaccines because I believe they are unsafe''. We found negative correlation with the perceived accuracy for the misleading Tweet with a warning tag (Figure 1a) and without a warning tag 
(Figure 1b) as shown in Table \ref{tab:H5}. The less participants were in favor of COVID-19 vaccines, the more accurate they perceived the misleading information regardless of the presence of a warning tag. We haven't found a significant correlation in the warning cover condition.

\begin{table}[!h]
    \centering
    \caption{Safety and Perceived Accuracy Tests: H5a/b}
    \label{tab:H5}
 \begin{tabularx}{\linewidth}{|l|Y|Y|}
  \hline
        & \textbf{r-test} & \textbf{Significance}  \\
  \hline 
  \textbf{H5a - with a warning tag} & $r = -.612$ & $p = .000^{*}$  \\\hline 
  \textbf{H5a - without a warning tag} & $r = -.329$ & $p = .017^{*}$ \\\hline
  \textbf{H5b - with a warning tag} & $r = -.344$ & $p = .011^{*}$ \\\hline
  \textbf{H5b - with a warning cover} & $r = -.473$ & $p = .000^{*}$ \\\hline
  \multicolumn{3}{|l|}{\textit{Significance Level: $\alpha = 0.05$}} \\\hline
 \end{tabularx}
\end{table}

The test of the same relationship for the case of the verified Tweet also revealed a negative correlation with the Tweet's perceived accuracy in both the warning tag (Figure 3b) and the warning cover condition (Figure 4). The more participants were in favor of COVID-19 vaccines the less accurate they perceived the verified information regardless if there was a warning tag or a warning cover. On a first thought, this might be a surprising result, but a careful consideration indicates a presence of a belief echo and resistance to \textit{soft moderated} verified information from \cite{Thorson}, but not the standalone verified Twitter content.

Next, we asked the participants to what extent do they agree with the following statement: ``There is no need to vaccinate for COVID-19 because I believe a natural herd immunity exists.'' We found negative correlation with the perceived accuracy for the misleading Tweet with a warning tag (Figure 1a) and without a warning tag (Figure 1b) as shown in Table \ref{tab:H6}. The more participants were in favor of the COVID-19 herd immunity, the more accurate they perceived the misleading information regardless if there was or there wasn't a warning tag. The test of the verified Tweet also revealed a negative correlation with the Tweet's perceived accuracy only in the original, no warning labels condition (Figure 3a). The more the participants were in favor of COVID-19 herd immunity the more accurate they perceived the verified information when no ``soft moderation'' intervention was applied. This finding adds to the evidence on the existence of echo beliefs in the essential need of COVID-19 vaccines as a preferred way of immunization.

\begin{table}[!h]
    \centering
    \caption{Herd Immunity and Perceived Accuracy Tests: H6a/b}
    \label{tab:H6}
 \begin{tabularx}{\linewidth}{|l|Y|Y|}
  \hline
        & \textbf{r-test} & \textbf{Significance}  \\
  \hline 
  \textbf{H6a - with a warning tag} & $r = -.529$ & $p = .000^{*}$  \\\hline 
  \textbf{H6a - without a warning tag} & $r = -.387$ & $p = .005^{*}$ \\\hline
  \textbf{H6b - without a  warning tag} & $r = -.445$ & $p = .001^{*}$ \\\hline
  \multicolumn{3}{|l|}{\textit{Significance Level: $\alpha = 0.05$}} \\\hline
 \end{tabularx}
\end{table}

\subsection{Belief Echoes: Efficacy and Hesitancy}
To test the subjective attitudes towards COVID-19 immunization, we asked the participants ``will it be possible to produce efficacious COVID-19 vaccines''. We found a significant result for the misleading Tweet with a warning tag  (Figure 1a) and with a warning cover (Figure 2) as shown in Table \ref{tab:H7}. In both cases, the participants who didn't believe in efficacious COVID-19 vaccines perceived the misleading Tweet ``somewhat acurate'' while the participants that did believe perceived it as ``not very accurate.'' We found a significant result for the verified Tweet in its original form, without any ``soft moderation'' (Figure 3a). In this case, the participants that didn't believe in efficacious COVID-19 vaccines perceived the original, non-moderated Tweet ``not very accurate.'' while the ones that did believe in the efficacy perceived it as ``somewhat accurate.''

\begin{table}[!h]
    \centering
    \caption{Efficacy and Perceived Accuracy Tests: H7a/b}
    \label{tab:H7}
 \begin{tabularx}{\linewidth}{|Y|Y|Y|}
  \hline
        & \textbf{r-test} & \textbf{Significance}  \\
  \hline 
  \textbf{H7a - with a warning tag} & $\chi^{2}(2) = 8.566$ & $p = .003^{*}$  \\\hline 
  \textbf{H7a - with a warning cover} & $\chi^{2}(2)  = 9.237$ & $p = .002^{*}$ \\\hline
  \textbf{H6b - without a warning tag} & $\chi^{2}(2) = 3.969$ & $p = .005^{*}$ \\\hline
  \multicolumn{3}{|l|}{\textit{Significance Level: $\alpha = 0.05$}} \\\hline
 \end{tabularx}
\end{table}

To test the personal hesitancy to COVID-19 immunization, we asked the participants ``Will you get vaccinated, if possible?'' We found a significant result for the misleading Tweet with a warning tag  (Figure 1a) and with a warning cover (Figure 2) as shown in Figure \ref{tab:H8}. In both cases, the participants that were hesitant to receive the COVID-19 vaccine perceived the misleading Tweet ``somewhat accurate'' while the participants that want to receive the vaccine as ``not very accurate.'' The participants that were unsure, perceived the misleading Tweet in both cases as ``not at all accurate.''. We found a significant result for the verified Tweet in its original form, without any ``soft moderation''  (Figure 3a). In the case, the participants that didn't want to receive the COVID-19 vaccine perceived the original, non-moderated Tweet ``somewhat accurate.'' while the participants that wanted to receive the vaccine perceived it as ``not very accurate.'' Similarly, the participants that were unsure perceived the Tweet as ``not at all accurate.''

\begin{table}[!h]
    \centering
    \caption{Personal Hesitancy and Perceived Accuracy Tests: H8a/b}
    \label{tab:H8}
 \begin{tabularx}{\linewidth}{|Y|Y|Y|}
  \hline
        & \textbf{r-test} & \textbf{Significance}  \\
  \hline 
  \textbf{H8a - with a warning tag} & $\chi^{2}(2) = 9.381$ & $p = .009^{*}$  \\\hline 
  \textbf{H8a - with a warning cover} & $\chi^{2}(2)  = 7.163$ & $p = .028^{*}$ \\\hline
  \textbf{H8b - without a warning tag} & $\chi^{2}(2) = 13.513$ & $p = .001^{*}$ \\\hline
  \multicolumn{3}{|l|}{\textit{Significance Level: $\alpha = 0.05$}} \\\hline
 \end{tabularx}
\end{table}

To test the hesitancy to COVID-19 immunization for children, we asked the participants ``Should children be vaccinated for COVID-19 too?'' We found a significant result for the misleading Tweet with only the warning tag  (Figure 1a) as shown in Table \ref{tab:H9}. The participants that were hesitant to administer the COVID-19 vaccine to children perceived the misleading Tweet with a warning tag ``somewhat accurate'' while the participants that agreed with administering the COVID-19 vaccine to children ``not very accurate.'' We found a significant result for the verified Tweet in its original form, without any ``soft moderation'' (Figure 3a). In the case, the participants that were hesitant to administer the COVID-19 to children perceived the an original, non-moderated Tweet as ``somewhat accurate.'' while the participants that agreed to administer the vaccine to children perceived it as ``not very accurate.''

\begin{table}[H]
    \centering
    \caption{Hesitancy for Children and Perceived Accuracy Tests: H9a/b}
    \label{tab:H9}
 \begin{tabularx}{\linewidth}{|Y|Y|Y|}
  \hline
        & \textbf{r-test} & \textbf{Significance}  \\
  \hline 
  \textbf{H9a - with a warning tag} & $\chi^{2}(2) = 10.663$ & $p = .001^{*}$  \\\hline 
  \textbf{H9b - without a warning tag} & $\chi^{2}(2) = 8.001$ & $p = .005^{*}$ \\\hline
  \multicolumn{3}{|l|}{\textit{Significance Level: $\alpha = 0.05$}} \\\hline
 \end{tabularx}
\end{table}

\subsection{Belief Echoes and Political Affiliation}
Following the association between one's political affiliation and the warnings of a misleading Twitter content \cite{Zannettou}, we analyzed the perceived accuracy among the participants based on their political affiliation (Republican, Democrat, independent). We found a significant difference in perception between the political affiliations of the participants for the misleading Tweet with the warning tag and without the warning tag as shown in Table \ref{tab:RQ1}. In both cases, the Republicans and independent participants perceived the Tweet as ``somewhat accurate'' while the Democrats as ``not very accurate.'' 

\begin{table}[H]
    \centering
    \caption{Political Affiliation and Perceived Accuracy Tests: RQ1}
    \label{tab:RQ1}
 \begin{tabularx}{\linewidth}{|Y|Y|Y|}
  \hline
        & \textbf{r-test} & \textbf{Significance}  \\
  \hline 
  \textbf{Misleading Tweet with a warning tag} & $\chi^{2}(2) = 7.063$ & $p = .029^{*}$  \\\hline 
  \textbf{Misleading Tweet without a warning tag} & $\chi^{2}(2) = 9.127$ & $p = .005^{*}$ \\\hline
  \multicolumn{3}{|l|}{\textit{Significance Level: $\alpha = 0.05$}} \\\hline
 \end{tabularx}
\end{table}

We also analyzed the association between political affiliation, the beliefs of safety and herd immunity, and the subjective attitudes on the vaccine efficacy and hesitancy. While there were no significant correlations about the safety and herd immunity beliefs, we found significant differences on the question of producing efficacious vaccines, personal hesitancy, as shown in Table \ref{tab:RQ2}. Almost one in four Republicans and and one in six Independents don't expect to have an efficacious COVID-19 vaccine (Table \ref{table_1}), while that proportion for the Democrats is one in forty. Half the Republicans and a third of the Independents are hesitant to receive the COVID-19 vaccine (Table \ref{table_2}), while only a tenth of the Democrats won't proceed with personal immunization. Roughly 40\% of the Republicans and Independents are hesitant to vaccinate children for COVID-19, to which only 8.3\% of the Democrats agree with (Table \ref{table_3}).  

\begin{table}[H]
    \centering
    \caption{Political Affiliation, Beliefs, and Subjective Attitudes Tests: RQ2}
    \label{tab:RQ2}
 \begin{tabularx}{\linewidth}{|Y|Y|Y|}
  \hline
        & \textbf{r-test} & \textbf{Significance}  \\
  \hline 
  \textbf{Producing efficacious vaccines} & $\chi(1) = 22.059$ & $p = .001^{*}$  \\\hline 
  \textbf{Personal Hesitancy} & $\chi(2)(2) = 55.486$ & $p = .000^{*}$ \\\hline
  \textbf{Hesitancy for Children} & $\chi(1) = 45.665$ & $p = .000^{*}$ \\\hline
  \multicolumn{3}{|l|}{\textit{Significance Level: $\alpha = 0.05$}} \\\hline
 \end{tabularx}
\end{table}

\begin{table}[h]
\renewcommand{\arraystretch}{1.2}
\caption{Pearson Chi-Square Test - Political Affiliation vs Production of an Efficacious Vaccine}
\label{table_1}
\centering
\begin{tabularx}{\linewidth}{|l|Y|Y|Y|}
\hline
& \textbf{Republicans} & \textbf{Democrats} & \textbf{Independent} \\ 
\hline
\textbf{Agree} & 46 (78.0\%) & 153 (97.5\%) & 86 (89.6\%)  \\
\hline
\textbf{Disagree} & 13 (22.0\%) & 4 (2.5\%) & 16 (15.7\%) \\
\hline
\end{tabularx}
\end{table}

\begin{table}[!h]
\renewcommand{\arraystretch}{1.2}
\caption{Pearson Chi-Square Test - Political Affiliation vs Personal Vaccination Hesitancy}
\label{table_2}
\centering
\begin{tabularx}{\linewidth}{|l|Y|Y|Y|}
\hline
& \textbf{Republicans} & \textbf{Democrats} & \textbf{Independent} \\ 
\hline
\textbf{Certain} & 28 (47.5\%) & 136 (86.6\%) & 63 (61.8\%)  \\
\hline
\textbf{Hesitant} & 27 (45.8\%) & 8 (5.1\%) & 35 (34.3\%) \\
\hline
\textbf{Undecided} & 4 (6.8\%) & 13 (8.3\%) & 4 (3.9\%) \\
\hline
\end{tabularx}
\end{table}

\begin{table}[!h]
\renewcommand{\arraystretch}{1.2}
\caption{Pearson Chi-Square Test - Political Affiliation vs Children Vaccination Hesitancy}
\label{table_3}
\centering
\begin{tabularx}{\linewidth}{|l|Y|Y|Y|}
\hline
& \textbf{Republicans} & \textbf{Democrats} & \textbf{Independent} \\ 
\hline
\textbf{Certain} & 34 (57.6\%) & 144 (91.7\%) & 61 (59.8\%)  \\
\hline
\textbf{Hesitant} & 25 (42.4\%) & 13 (8.3\%) & 41 (40.2\%) \\
\hline
\end{tabularx}
\end{table}

\section{Discussion}
Consistent with the previous evidence on receptivity to misinformation and resistance to warnings \cite{Clayton, Nyhan}, we found that the more likely participants were to believe that COVID-19 vaccines are unsafe, the more receptive they were to misleading COVID-19 vaccines information from Twitter, resisting the soft moderation intervention, proving the existence of belief echoes. 

\subsection{Strength and Type of Warning Label}

That the participants perceived the misleading Tweets as accurate in the presence of a warning tag but not in the presence of a cover condition suggests that the warning tags are not effective or insufficient to sway participants' perceived accuracy. These findings are consistent with previous research showing that the design of the warning label affects individuals' perceptions of the content \cite{Kaiser, Seo, Moravec, Clayton}. The design of warning labels and how they are presented to users can impact the warning labels’ effectiveness, with more explicit labels being more effective \cite{Moravec}. For instance, it was found that individuals routinely ignored contextual warnings, which were akin to the warning tags in our study in that they did not obscure the misleading content nor require individuals to click through to see the information \cite{Kaiser}.

However, interstitial warnings, akin to cover warnings in our study in that they required individuals to click through to continue, were effective in countering disinformation. Authors in \cite{Kaiser} posit that this may be because interstitial designs are more noticeable for users and thereby more effective at countering misinformation. It may also be that these designs require users' engagement and thus necessitate a cognitive awareness of the tag’s content. Similarly, authors in \cite{Clayton} found that the perceived accuracy decreased with the increasing strength of the warning tags, such that tags which said ``rated false'' were significantly more effective than ``disputed'' tags at reducing beliefs in the misleading information. These findings suggest that more explicit, unambiguous warnings are more effective at countering misleading information. In our case, the warning covers - not the tags, which are more verbose and exact, were a more potent way of urging users to critically discern COVID-19 vaccine content on Twitter. However, users may habituate to the cover warnings in the long-term if they perceive that the moderator, Twitter, is biased in labeling content from users with particular political leanings \cite{Burrell}.

\subsection{Preconceived Notions-Explaining Results of Verified Tweets}
Dispelling belief echoes on Twitter might be a more complex task, dependent on the content or type of misinformation and the subjective involvement of the participants. The fact that participants who were more likely to believe that COVID-19 vaccines were unsafe, were less likely to perceive the verified Tweet as inaccurate in both the warning tag and cover conditions may be due to the fact that the verified Tweet, though accurate, still reflected information that was negative about COVID-19 vaccines (i.e., invoking the idea that they may lead to serious side-effects and should be avoided by some participants in some situations). In other words, it would make sense that participants favouring vaccines would be more likely to disbelieve the Tweet expressing a concern about COVID-19 vaccines when accompanied by a warning label. A similar conclusion could be drawn also in the case of belief in herd immunity versus mass immunization with COVID-19 vaccines. 

\subsection{COVID-19 Vaccine Beliefs}
In terms of efficacy, participants who thought that the COVID-19 vaccines were ineffective were more likely to rate the misleading Tweets as accurate, regardless of the soft moderation applied. This suggests that belief echoes persist despite the warnings, and perhaps, a presence of a warning tag or cover may actually increase people's likelihood of finding a misleading Tweet accurate if it conforms to existing beliefs. This finding is consistent with the backfire effect previously observed for polarizing content on social media \cite{Pennycook}. For example, evidence suggests that corrections on misleading Tweets strengthened misperceptions (or perceptions of accuracy) among those most strongly committed to the belief \cite{Nyhan}. The corrections that contradict users’ preconceived notions were found to lead individuals to double down on their beliefs. 

In terms of hesitancy (both personal and for children vaccination), we found a similar effect, such that those who were hesitant about vaccines were more likely to perceive misleading Tweets with tags and covers as accurate while those who wanted the vaccine perceived it as inaccurate. Again, the fact that only Tweets with tags and covers were viewed as accurate suggests evidence for a backfire effect such that the mere presence of the tags/covers may increase individuals’ beliefs in the Tweets' accuracy if the content reinforces the participants’ anti-vax stance. A similar conclusion follows from the fact that, for the original (verified) Tweet, the pro-vax participants who wanted to get the vaccine viewed it as not very accurate but the anti-vax participants viewed it as accurate.

\subsection{Political Affiliations}
Along the lines of the findings in \cite{Zannettou, Christenson, Nyhan}, we found further evidence of the association between user's political affiliations and the receptivity to misleading content. The Republican and Independent participants perceived the Tweet as ``somewhat accurate'' while the Democrat participants perceived it as ``not very accurate'' in both the misleading Tweet with and without a warning tag. That the difference between the expectation of an efficacious COVID-19 vaccine is twentyfold between the Republicans and Democrats is a bit surprising, but consistent with the breakdown of trust in scientists to deliver an efficacious COVID-19 vaccine along the party line \cite{Funk}. The hesitancy we found in our study is consistent with the previous reported breakdown for the COVID-19 vaccine hesitant Republicans and Democrats, both personally and in regards to children's vaccination \cite{Karson}. Interestingly, Independents showed a high hesitancy on par with the Republicans in both cases. 


Authors in \cite{Christenson} found that while corrections had the intended effect among Democrats, soft moderation techniques backfired among Republicans. Specifically, the authors found that while corrections of misleading claims decreased Democrats’ perceptions of claim accuracy, they actually strengthened Republicans’ perceptions of accuracy. As in \cite{Nyhan}, these findings suggest that corrections of misleading information on social media may not only be ineffective among some individuals but may actually reinforce individuals’ preconceived notions. While our study did not assess participants’ beliefs before and after receiving corrections, as all participants were only assessed once, the findings that political affiliation affects individuals’ perceptions of accuracy and the impact that warning labels have on those perceptions are consistent with the backfiring effect among individuals with certain political ideologies.

\subsection{Usable Security and Privacy Implications}
Reluctance to heed security or privacy warnings is not a new phenomenon and has been well researched in the past \cite{Garfinkel, Nicholson, Fagan}. While efforts have been invested in increasing the clarity of the messages and design of affordances to attract attention and motivate users, habituation is a complex problem transcending security designs. Habituation describes a diminished response with repetitions of the same stimulus, decreasing the intended effect of security and privacy warnings among users. Authors in \cite{Vance}, in this context, have uncovered the phenomenon of ``generalization'' where habituation to one stimulus carries over to other novel stimuli that are similar in appearance. We didn't explore the diminished response with repetitions of the same warning label to a Tweet, being that a tag or a cover, but that certainly warrants close research attention. The findings of our study suggest that heeding a misleading information warning only happened when the information is obscured by a plain text warning of the risks, not when the warning follows the Tweet with tag. 

The warning tag, consisting of an exclamation mark symbol urges users to ``Get the facts about COVID-19,' in Twitter's blue font, communicates a seemingly ambiguous message without explicitly addressing that the Tweet's content aims to mislead users about the COVID-19 vaccine. Perhaps a line of research could explore a variation of tags that are more direct, for example ``This is COVID-19 misinformation'', written in bold red font and conventional warning favicons. Alternatively an impartial message like ``No judgment, but this might be COVID-19 misinformation'' could also show users' receptivity to not-so-overt moderation. The warning cover, along these lines, communicated a message where Twitter appeared not taking sides by saying: \textit{This Tweet violated the Twitter Rules about spreading misleading and potentially harmful information related to COVID-19. Twitter has determined that may be in the public interest for the Tweet to Remain Accessible.} It also provided a link for the participant to \textit{Learn More}, which largely subsumes the warning tag by leading users to a repository of verified COVID-19 information. Alternative wording like \textit{This Tweet was rated `false', but we keep it in the public interest}, based on the previous evidence \cite{Clayton}, could yield a stronger reduction of misperception.

In the context of generalization, Twitter labels alternative narratives that are: (1) statements or assertions that have been confirmed to be false or misleading by subject-matter experts, such as public health authorities (\textit{misleading information}); (2) statements or assertions in which the accuracy, truthfulness, or credibility of the claim is contested or unknown (\textit{disputed claims}); and (3) information (which could be true or false) that is unconfirmed at the time it is shared (\textit{unverified claims})\cite{Roth}. With a similarity in labeling alternative narratives, from a user perspective, habituation to a tag or a cover for misleading COVID-19 vaccination could potentially carry over to other warnings about other polarizing events, such as elections. An interesting line of research could investigate the generalization effect not just between different labels, but in various combinations of formatting and wording. A user might be well aware and agree that some claims about elections are disputed, but they can nonetheless retain their beliefs about the COVID-19 vaccine safety and efficacy. Further so, another line of research could be an investigation of the generalization phenomenon between social media platforms, for example between warning labels on Twitter and Facebook.  

This discussion brings an important aspects of usable security affordances that depart from the conventional warning about system-level exploits towards content-level warnings. System-level exploits hardly relate to any potent beliefs (outside perhaps of the stereotypical foreign nation-state interference) or better said, users might not have strong polarizing stances on phishing or malware, usually perceiving it as a ``bad thing'' \cite{Felt}. Content-level exploits, on the other side, are far more complex and potent in polarizing users, given that they are subjectively involved with the content \cite{Stewart}. Users might ignore a red screen proceeding to a suspicious website, but they usually trust Chrome or Firefox that they have honest intentions in warning them about potential risks. Evidence already indicates that users are not trusting of the soft moderation intervention, feeling that Twitter itself was biased and mislabeling content \cite {Geeng}. Remaining impartial while trying to dispel belief echoes might be a harder problem depending on the content - while there are safe and unsafe websites, there is, and will continue to be, a wealth of polarizing content on Twitter that will require content-relevant warning labeling.   

\subsection{Ethical Implications}
While we set out to investigate the effect of soft moderation on Twitter and debriefed the participants at the end of the study, the results could have several ethical implications, nonetheless. We exposed the participants to a misleading and manipulated soft moderation of twitter content on the topic of the COVID-19 vaccine that could potentially affect participants' stance on vaccination and the pandemic. The exposure might not sway participants on the hesitancy or their perceptions of safety and efficacy, but could make the participants reconsider their approach of obtaining the vaccine for themselves or their children. The exposure could also affect the participants' stance of social media soft moderation in general and nudge people to move to less regulated platforms \cite{Caulfield}. A recent example of such a migration from Twitter to Parler, Rumble and Newsmax was witnessed after Twitter actively labeled and removed false information on the platform during the 2020 U.S. elections ~\cite{Isaac}. 

That the participants were able to critically discern the content of the verified Tweet despite our alternation to include warning labels is reassuring and suggests that misinformation has the potential to be contained, if not eradicated, from social media platforms. However, the potential of crafting software that could silently attach or remove warning covers before they are presented to Twitter users could have unintended consequences. In the past, such an effort was tested in manipulating a Twitter textual content (not any additional affordances in the user interface) to induce misperceptions about the relationship between vaccines and autism \cite{twittermim}. With the evidence of nation-states censoring Twitter regarding narratives countering their interest in the past, it is possible that such a nation-state could use a similar approach and implement a ``post-soft moderation'' logic within a state-approved and disseminated social media application \cite{Thomas}. This may be far from the realm of possibility, even if the capabilities exist, but for such a sensitive topic as COVID-19 vaccination, meddling with the warning labels could give an edge to a vaccine competitor in the global race for development and procurement of COVID-19 vaccines. We condemn such ideas and use of our research results. Evidence for such a nefarious misinformation Twitter campaign that promotes homegrown Russian vaccine and undercuts rivals has already surfaced \cite{Frenkel}.

Perhaps outside the scope of this study, the ethical questions remains whether Twitter, or any social media platform, acting as a private entity, could set a precedent of an ultimate arbiter of what constitutes misinformation and what does not. Twitter most likely applies an automated means of warning labeling in conjunction with manual moderation, as evidence with the strange labeling of Tweets that contained the words ``oxygen'' and ``frequency'' for COVID-19 related Tweets \cite{Zannettou}. Even with an attempt at honest moderation, cross-checked with the health authorities like CDC, a potential problem might arise in case a previously held belief, or a fact about COVID-19 is later disputed. For example, at the beginning of the pandemic, authorities claimed masks were not effective in protecting the virus from spreading, a claim that later was not reversed, resulting in masks becoming essential to any human-to-human interaction \cite{Zhang}. So if the warning labels were applied to moderate any Tweet that contains the words ``mask'' and ``stop'' or ``spread'' at the early periods of the pandemic, they must be retracted. Such a thing could cast doubt on studies like ours, even if we as researchers, and Twitter as moderators, acted in good faith. Certainly, this could damage the reputation of users as well as Twitter and further exacerbate the impression of not-so-honest impartiality in labeling content as misleading, especially against users identifying themselves as conservatives \cite{Burrell}.   

\subsection{Future Research}
We acknowledge that there is further research to be done into investigating the full ramifications of soft moderation by social media platforms, especially beyond the topics of the COVID-19 pandemic or presidential elections. A promising line of research is the combination of soft and hard moderation, given that Twitter has exercised the right to ban or suspend accounts indefinitely that have been labeled for misinformation in the past, like in case of Donald Trump. One could probe the warning labeling algorithm and reverse engineer it to find if there is a relationship between the number/type of warning labels an account could receive, for example, before it gets permanently banned (if an automated ban exists, given the wide latitude and the imperative of Twitter to remain not overly controlling of the public discourse). When social media platforms, as private entities, are predetermined by users to hold biased positions \cite{Geeng}, any action taken to apply soft moderation techniques may be undermined in the process, instead working to legitimize the beliefs of skeptics. In this direction, a content analysis of the Tweets being labeled could reveal the topics, images, words, or the network of accounts behind such impressions.  

In the United States, where the right to speech is protected, more research may be done to see how alternative narratives, belonging to a same type of content or a topic (e.g. COVID-19 vaccines cause adverse effects leading to death)  are soft moderated between platforms, for example comparing Twitter, Facebook, or Parler. Soft moderated content is usually closely related to content used for trolling so this relationship could be also explored, such as understanding if warning labeled Tweets provoke emotional response and of what kind. Similarly, the warning labeling can be associated to identify the evolution of political information operations on Twitter, that have been waged on the topic of COVID-19 already ~\cite{Strick}.


\subsection{Scope Limitations}
The current study has important limitations. First, it is possible that a different topic or even different information regarding the effect of the COVID-19 vaccines would have different outcomes. We used a couple of Tweets that were relevant to the state of the pandemic and mass immunization during the period of January-February 2021, which could be perceived with a different level of accuracy after a certain period of time. We used only two Tweets and a study that explores the effect on multiple misleading or verified Tweets on COVID-19 vaccine could uncover different efficiency or strength of soft moderation. Overall, the findings in the present study may be specific to the effect that warning tags have on COVID-19 (mis)information and cannot be generalized to other topics. Second, participants who are more regular social media users in general may be desensitized to the information presented in the Tweets, which may have affected their perception of the issue irrespective of the warnings. 

Third, our experiment was limited to Twitter as a social media platform of choice. Since the content and images we present are borrowed and adapted to the study objectives from Twitter, we are limited to evaluating the effects of the warning tags and warning cover on perceptions of accuracy on Twitter only and may not be generalized regarding other social media platforms. We were limited to the formatting and wording of the warning labels chosen by Twitter at the time of the study. If Twitter chooses to place the tag, say on top of the Tweet instead of the bottom, the results could be different. Similarly,  if the wording of the warning cover changes, the results might not hold for such new conditions. Fourth, we did not examine the effects over a period of time. Thus, we are unable to examine the Tweet’s effects following the study. We also acknowledge another limitation imposed of the timeline of the study and the speed of COVID-19 vaccine development. By the time participants completed the study, much more might be known about the COVID-19 vaccines to sway public opinion. For instance, if many participants have gotten the vaccine without major side effects by the time participants complete the study, this might affect their responses. Fifth, although we tried to sample a representative set of participants for our study using Amazon Mechanical Turk, the outcomes might have been different if we used another platform, or other type of sampling. Also, a much larger sample size could have provided a more nuanced view of the soft moderation, but we had limited funding for this study.

\section{Conclusion}
In the present study, we sought to determine whether two forms of soft moderation on Twitter affect the perceived accuracy of Tweets pertaining to COVID-19 vaccines. We were also interested in examining whether perceived Tweet accuracy varies based on individual's' beliefs regrading COVID-19 vaccine safety, efficacy, willingness to receive vaccinations, and their political affiliations. Overall, our results suggest that warning covers are more effective than warning tags in dispelling individuals’ beliefs about misleading Tweets. Individuals’ pre-existing beliefs regarding COVID-19 vaccine safety, efficacy, and hesitancy affect individuals’ perceptions of Tweet accuracy such that their perceptions of the accuracy align with their biases. Furthermore, our results also show that individuals' political affiliations also affect their perceptions of accuracy for misleading Tweets such that Republicans and Independents, who are more likely to express skepticism regarding vaccines, are more likely to perceive misleading Tweets as accurate, irrespective of any moderation effort. 

In all cases, individuals perceive the Tweets in ways that are most favourable to or consistent with their pre-existing beliefs. This may lead to a backfire effect, as evidenced by the fact that individuals who were skeptical of vaccines were more likely to rate misleading Tweets with tags and covers, but not misleading Tweets without tags, as accurate. Taken together, our findings provide additional evidence for the existence of belief echoes pertaining to COVID-19 vaccines that are largely resistant to soft moderation in the form of warning tags but not warning covers. We believe that the insight gained from this research regarding how individuals’ pre-existing belief biases impact perceptions of Tweet accuracy in the context of soft moderation can be used to develop more effective moderation techniques that do minimize unintended consequences. We hope that our results could inform the usable security community towards future steps in eradicating misinformation on Twitter and social media in general.

\bibliographystyle{ACM-Reference-Format}
\bibliography{sm}

\section*{Appendix}\label{sec:appendix}
The study questionnaire included the following questions: 

\begin{itemize}
    \item \textbf{Perceived Accuracy of a Tweet:} \\1. \textit{To the best of your knowledge, how accurate is the claim described in the Tweet?} \\4-point Likert scale (1-not at all accurate, 2-not very accurate, 3-somewhat accurate, 4-very accurate).  
    
    \item \textbf{Beliefs:} \\2. \textit{How much do you agree with the following statement:''I am not favorable to vaccines because they are unsafe''?} \\3. \textit{How much do you agree with the following statement:''here is no need to vaccinate because a natural immunity exists''?} 
    \\4-point Likert scale (1 - Totally, 2 - A Little, 3 - Partially, 4 - Not at All). 
    
    \item \textbf{Subjective Attitudes:} \\4. \textit{Will it be possible to produce safe and efficacious COVID-19 vaccines?}\\ Yes/No. \\5. \textit{Will you get vaccinated, if possible?}\\ Yes/No/I Don't Know. \\6. \textit{Should children be vaccinated for COVID-19 too?}\\ Yes/No. 
    
    \item \textbf{Demographics}: \\ Age, gender identity, education, political leanings.
    
\end{itemize}

\end{document}